\documentclass{article}

  \usepackage[final, nonatbib]{arxiv2020}

\usepackage[utf8]{inputenc} % allow utf-8 input
\usepackage[T1]{fontenc}    % use 8-bit T1 fonts
\usepackage{hyperref}       % hyperlinks
\usepackage{url}            % simple URL typesetting
\usepackage{booktabs}       % professional-quality tables
\usepackage{amsfonts}       % blackboard math symbols
\usepackage{nicefrac}       % compact symbols for 1/2, etc.
\usepackage{microtype}      % microtypography
\usepackage{graphicx}

\usepackage{color}

\newcommand*\samethanks[1][\value{footnote}]{\footnotemark[#1]}

\title{Protein model quality assessment using rotation-equivariant, hierarchical neural networks}

\author{
  Stephan Eismann\thanks{Equal contribution}  \\
  Department of Applied Physics\\
  Stanford University\\
  \texttt{seismann@stanford.edu} \\
  \And
  Patricia Suriana\samethanks \\
  Department of Computer Science\\
  Stanford University\\
  \texttt{psuriana@stanford.edu} \\
   \AND
 Bowen Jing \\
  Department of Computer Science\\
  Stanford University\\
  \texttt{bjing@stanford.edu} \\
   \And
 Raphael J.L. Townshend \\
  Department of Computer Science\\
  Stanford University\\
  \texttt{raphael@cs.stanford.edu} \\
     \And
 Ron O. Dror \\
  Department of Computer Science\\
  Stanford University\\
  \texttt{rondror@cs.stanford.edu} \\
}

\begin{document}

\maketitle

\begin{abstract}
Proteins are miniature machines whose function depends on their three-dimensional (3D) structure. Determining this structure computationally remains an unsolved grand challenge. A major bottleneck involves selecting the most accurate structural model among a large pool of candidates, a task addressed in model quality assessment. Here, we present a novel deep learning approach to assess the quality of a protein model. Our network builds on a point-based representation of the atomic structure and rotation-equivariant convolutions at different levels of structural resolution. These combined aspects allow the network to learn end-to-end from entire protein structures. Our method achieves state-of-the-art results in scoring protein models submitted to recent rounds of CASP, a blind prediction community experiment. Particularly striking is that our method does not use physics-inspired energy terms and does not rely on the availability of additional information (beyond the atomic structure of the individual protein model), such as sequence alignments of multiple proteins.
\end{abstract}

\section{Introduction}
\label{sec:intro}

Proteins--—important components of the cell which perform a wide array of functions—--comprise long chains of amino acids that fold into compact globular 3D structures. Determination of this 3D structure is critical not only for understanding how proteins function, but also for designing drugs that can bind to a protein and alter its activities. Solving protein structures experimentally is difficult, time consuming and expensive, leading to the ever-increasing gap between available sequence data and available experimental structures. This gap amplifies the critical need for computational approaches that accurately predict protein structure from amino acid sequences.

Despite the recent advances in computational methods \cite{alphafold,itasser,intfold,multicom,quark}, protein structure prediction remains an unsolved grand challenge. This challenge generally involves two steps: sampling and scoring. Sampling describes the generation of candidate models of protein structure given a sequence. Scoring aims to select the best among the large pool of candidate models where best describes how close a given model is to the true structure. This latter task of model quality assessment has attracted the application of a number of deep-learning methods in recent years \cite{3dcnn,ornate,baldassarre2020graphqa,karasikov2019smooth,proq3d}.

Here, we introduce a deep-learning scoring function that assesses model quality given just the atomic coordinates and without the use of physics-inspired energy terms or other pre-computed features. Our method has several key characteristics: (1) equivariance to 3D rotations, which allows the network to recognize structural motifs independent of their orientation, (2) hierarchical layers that preserve rotation equivariance, allowing the network to identify structural motifs at many scales, (3) a focus on local interactions at each hierarchical level, reflecting the fact that inter-atomic forces are predominantly local, and (4) learning directly from atomic coordinates rather than mapping to a grid, allowing high spatial resolution even for large structures.

Our method shows state-of-the-art results in ranking protein models submitted to recent Critical Assessment of protein Structure Prediction (CASP) community experiments (CASP11-12) and does not rely on the availability of additional information, such as multiple sequence alignments.  

\section{Methods}

\subsection{Dataset}
We train and test our method on candidate models submitted to multiple rounds of CASP, a biennial community experiment. CASP \cite{kryshtafovych2019critical} addresses the protein structure prediction problem by withholding newly solved experimental structures (referred to as \textit{targets}) and allowing computational groups to make predictions (referred to as \textit{models}). Submitted models are released as sets in two stages (20 models per target for stage 1, 150 models per target for stage 2) for Model Quality Assessment (MQA), a specific subcategory of CASP that aims to assess the performance of scoring functions. Model quality is measured in terms of GDT\_TS \cite{zemla2001processing} based on the alignment of native structure and candidate model. GDT\_TS ranges between 0 and 1, with higher GDT\_TS value indicating better model quality.

We mirror the setup of the CASP experiment and split the CASP datasets based on protein target and release year. We train and validate our method on the set of models submitted to CASP 5-10 (500 targets for training, 58 for validation). For testing, we consider models submitted to stage 2 for CASP 11 (84 targets) and 12 (40 targets). 
We relaxed the structure of all models with the SCWRL4 software \cite{Dunbrack2009} to improve side-chain conformations prior to feeding the models into our network. 

\subsection{Architecture}
\begin{figure}[h]
\caption{\textbf{Network architecture} Given 3D coordinates and element types of every atom as input, the network performs rotation-equivariant convolutions over multiple layers. The network first learns features at the level of every atom before we aggregate information at the level of alpha carbons (a subset of all atoms) in the next layer. We subsequently average the features over all alpha carbons to obtain a fingerprint for the entire protein model. This fingerprint is the input to a shallow dense network that outputs a final scalar score.}
\centering
\vspace*{3mm}
\includegraphics[width=0.8\textwidth]{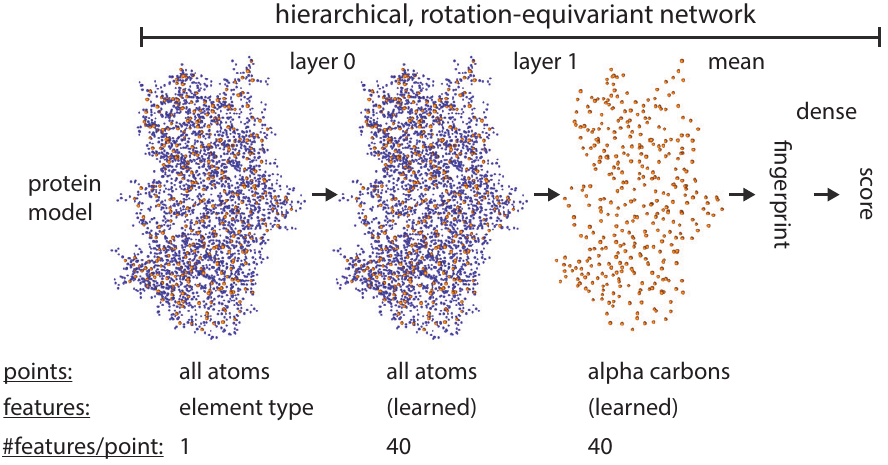}
\label{fig:schematic}
\end{figure}

Our network builds on recent neural network architectures that are specifically designed to learn from 3D atomic structures \cite{thomas2018tensor, eismann2020hierarchical}. Figure~\ref{fig:schematic} illustrates the architecture of our network. At its core are a point-based representation of atoms and multiple layers of rotation-equivariant convolutions. The goal is to predict a quality score for a given protein model.
Each atom has an associated feature vector. At input this is simply the one-hot encoding of its element type\footnote{We represent carbon, oxygen, nitrogen, and sulfur atoms.}. 
We then perform rotation-equivariant convolutions that result in a new feature vector associated with each atom. The convolution filters of each layer are constructed based on a truncated series of spherical harmonics (up to rotation order $l=2$), such that the filters are able to recognize structural motifs independent of their orientation or position in space. Convolutions are performed among a limited set of k-nearest neighbors (k=40) to account for the fact that the physical laws governing intra-molecular interactions are local. The subsequent convolution layer outputs features only for the alpha carbon of each amino acid residue. This subsampling operation aggregates information hierarchically and allows the network to recognize structural motifs at different scales. We average the alpha carbon feature vectors to obtain a fingerprint for the entire protein model\footnote{Instead, we could also choose to aggregate information at the level of a single point using further subsampling operations (see \cite{eismann2020hierarchical}). The optimal choice of hierarchy is likely application dependent.}. From this fingerprint, we use two dense layers (250 and 150 units, ReLu activation) to calculate a single scalar quality score.

\subsection{Training and Evaluation}
We formulate the training as a regression task aiming to predict the quality metric GDT\_TS for each model. The Huber loss between the actual and the predicted GDT\_TS is used as the loss function. We train with the Adam optimizer in TensorFlow \cite{tensorflow2015-whitepaper} (learning rate $1.25 \cdot 10^{-4}$) and monitor the loss on the validation set for every epoch. The weights of the best-performing network are then used to evaluate the predictions on the test set. We use Horovod \cite{sergeev2018horovod} to distribute training across 4 NVIDIA Titan X GPUs. 

\section{Results}
\subsection{Model quality assessment}
We examine our results on model quality assessment, finding that we generally improve upon state-of-the-art methods (Table~\ref{tab:casp11-12}). We report multiple correlation metrics per method. \emph{Global correlation} indicates the correlation between a method’s predictions and the GDT\_TS scores of all protein models in a given set. \emph{Per target} indicates the correlation  with respect to a method’s predictions and the GDT\_TS scores of the protein models for a given target (averaged over all targets). 
The two measures provide complementary information about a method's performance. Good global correlation is desirable to judge the absolute quality of a set of candidate models. Per target correlation indicates a method's ability to distinguish model quality among a set of models for one target (the main scoring challenge). We report Pearson, Kendall, and Spearman correlation coefficients for both global and per target correlations. 
Notably, our method also improves upon ProQ3D \cite{proq3d}, which uses information on related proteins to make its predictions\footnote{GraphQA can also leverage information on related proteins, but here we compare against the version of GraphQA that only uses structural information.}.

%% Three coefficients, stage 2
\begin{table}  
  \caption{Comparison with state-of-the-art methods on CASP 11 and 12 in terms of global and mean per-target correlation coefficients (higher is better). The different coefficients are Pearson (r), Spearman ($\rho$), and Kendall ($\tau$). The top performing method for each metric is shown in bold.}
  \label{tab:casp11-12}
    \centering
\begin{tabular}{@{}llccclccclccclcc@{}}
\toprule
                               & \multicolumn{7}{c}{CASP 11, stage 2}                                                             &  &               & \multicolumn{6}{c}{CASP 12, stage 2}                                             \\ \midrule
                               & \multicolumn{3}{c}{Global}                    &  & \multicolumn{3}{c}{Per target}                &  & \multicolumn{3}{c}{Global}                    &  & \multicolumn{3}{c}{Per target}                \\ \cmidrule(lr){2-4} \cmidrule(lr){6-8} \cmidrule(lr){10-12} \cmidrule(l){14-16} 
Method                         & r             & $\rho$           & $\tau$           &  & r             & $\rho$           & $\tau$           &  & r             & $\rho$           & $\tau$           &  & r             & $\rho$           & $\tau$           \\ \midrule
Ours                           & \textbf{0.84} & \textbf{0.84} & \textbf{0.65} &  & \textbf{0.45} & \textbf{0.43} & \textbf{0.31} &  & 0.80 & 0.79          & 0.59          &  & \textbf{0.62} & \textbf{0.55} & 0.39 \\
3DCNN \cite{3dcnn}                    & 0.64          & 0.69          & 0.48          &  & 0.40          & 0.39          & 0.27          &  & 0.61          & 0.64          & 0.46          &  & 0.51          & 0.45          & 0.32          \\
Ornate \cite{ornate}                  & 0.63          & 0.67          & 0.48          &  & 0.39          & 0.37          & 0.26          &  & 0.67          & 0.66          & 0.47          &  & 0.49          & 0.46          & 0.32          \\
GraphQA \cite{baldassarre2020graphqa} & 0.82          & 0.82          & 0.62          &  & 0.38          & 0.36          & 0.25          &  & \textbf{0.81} & \textbf{0.81} & \textbf{0.62} &  & 0.61          & \textbf{0.55}          & \textbf{0.40}          \\
VoroMQA \cite{olechnovivc2017voromqa} & 0.65          & 0.69          & 0.51          &  & 0.42          & 0.41          & 0.29          &  & 0.61          & 0.60          & 0.45          &  & 0.56          & 0.50          & 0.36          \\
SBROD \cite{karasikov2019smooth}      & 0.55          & 0.57          & 0.39          &  & 0.43          & 0.41          & 0.29          &  & 0.47          & 0.49          & 0.34          &  & 0.61          & \textbf{0.55}          & \textbf{0.40}          \\
ProQ3D \cite{proq3d}                  & 0.77          & 0.80          & 0.59          &  & 0.44          & \textbf{0.43}          & 0.30          &  & \textbf{0.81} & 0.80          & 0.60          &  & 0.60          & 0.54          & 0.39          \\ \bottomrule
\end{tabular}
\end{table}

\subsection{Visualization of learned embeddings}
\begin{figure}[h]
\centering
\includegraphics[width=0.72\textwidth]{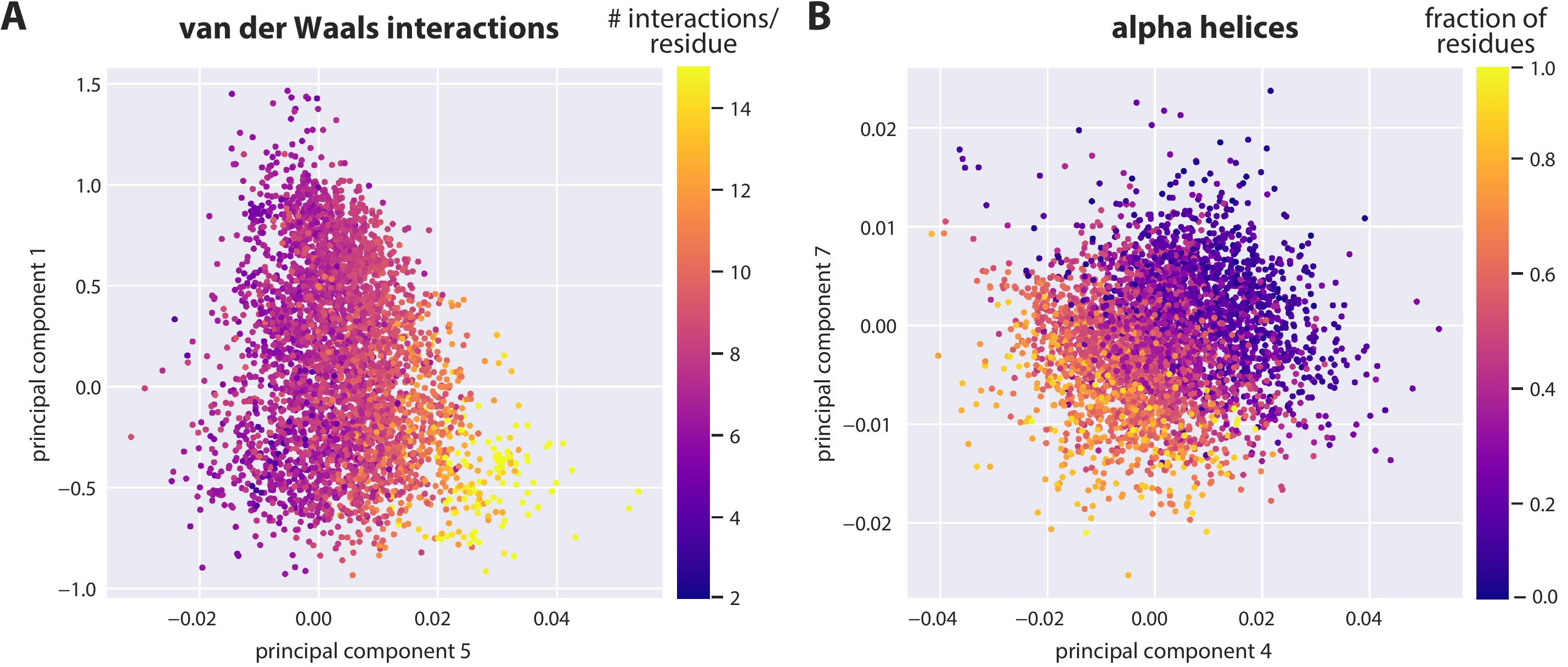}
\caption{\textbf{PCA projections of learned fingerprints}. PCA projections of the protein model fingerprints reveal their encoding of structural motifs. (A) shows clustering of the protein models based on the number of van der Waals interactions per residue. All models with 15 or more interactions per residue are shown in the same bright yellow color. (B) shows the fraction of residues within each protein model that is part of an alpha helix.}
\label{fig:pca}
\end{figure}

In Figure \ref{fig:pca}, we explore whether the network has learned to encode certain structural motifs. We project the fingerprint (see Figure \ref{fig:schematic}) of each protein model in the test set  into a lower-dimensional space using Principal Component Analysis (PCA). Visual inspection reveals that the fingerprint contains information on both inter-atomic interactions and secondary protein structure. 

In Figure \ref{fig:pca}A, we consider the prevalence of van der Waals interactions in each protein model. We use the software GetContacts \cite{getcontacts} to identify van der Waals interactions and divide the total number of interactions by the number of amino acid residues per model. We note that protein models cluster based on the prevalence of van der Waals interactions in the plane of principal component (PC) 5 and PC 1.
In Figure \ref{fig:pca}B, we consider the prevalence of alpha helices in a given protein model. We use DSSP \cite{dssp} to assign each residue in a protein model to one type of secondary structure ('alpha helix', 'beta sheet' or 'other'). We then calculate the fraction of 'alpha helix' residues over all residues in a model. Models cluster based on this fraction when plotted based on PC 4 and PC 7.  

\section{Discussion}
In this work, we presented a hierarchical deep-learning method to assess the quality of candidate models of protein structure. Our method learns end-to-end from the 3D atomic coordinates of protein models without the use of any physics-inspired or statistical energy terms. Thanks to the rotation equivariance of the network filters, the orientation in which motifs and models are presented to the network does not matter.

Our results on the CASP datasets indicate improved global and per-target GDT$\_$TS correlations compared to previous approaches, including approaches that use additional information such as multiple-sequence alignments. 

The fact that our network learns to predict a quality score given solely the atomic coordinates of a single protein model makes it suitable to guide sampling in protein structure modelling algorithms such as Rosetta \cite{rosetta}, which we leave as future work.

\clearpage
\bibliographystyle{main_arxiv}
\bibliography{main_arxiv}

\end{document}